# GEFF: Graph Embedding for Functional Fingerprinting

**Kausar Abbas[1,2], Enrico Amico[1,2], Diana Otero Svaldi[3,4], Uttara Tipnis[1,2], Duy Anh Duong-Tran[1,2], Mintao Liu[1,2], Meenusree Rajapandian[1,2], Jaroslaw Harezlak[5], Beau M. Ances[6], Joaquín Goñi[1,2,7]**

[1] Purdue Institute for Integrative Neuroscience, Purdue University, West Lafayette, IN, USA
[2] School of Industrial Engineering, Purdue University, West Lafayette, IN, USA
[3] Indiana University School of Medicine, Indiana University, Indianapolis, IN, USA
[4] Indiana Alzheimer Disease Center, Indiana University, Indianapolis, IN, USA
[5] Department of Epidemiology and Biostatistics, Indiana University, IN, USA
[6] Washington University School of Medicine, Washington University, St Louis, MO, USA
[7] Weldon School of Biomedical Engineering, Purdue University, West Lafayette, IN, USA

**ABSTRACT**

It has been well established that Functional Connectomes (FCs), as estimated from functional MRI (fMRI) data, have an individual fingerprint that can be used to identify an individual from a population (*subject-identification*). Although identification rate is high when using resting-state FCs, other tasks show moderate to low values. Furthermore, identification rate is task-dependent, and is low when distinct cognitive states, as captured by different fMRI tasks, are compared. Here we propose an embedding framework, GEFF (Graph Embedding for Functional Fingerprinting), based on group-level decomposition of FCs into eigenvectors. GEFF creates an eigenspace representation of a group of subjects using one or more task FCs (*Learning Stage*). In the *Identification Stage*, we compare new instances of FCs from the *Learning* subjects within this eigenspace (validation dataset). The *validation* dataset contains FCs either from the same tasks as the *Learning* dataset or from the remaining tasks that were not included in *Learning*. Assessment of validation FCs within the eigenspace results in significantly increased *subject-identification* rates for all fMRI tasks tested and potentially *task-independent* fingerprinting process. It is noteworthy that combining resting-state with *one* fMRI task for GEFF *Learning Stage* covers most of the cognitive space for subject identification. Thus, while designing an experiment, one could choose a task fMRI to ask a specific question and combine it with resting-state fMRI to extract maximum subject differentiability using GEFF. In addition to subject-identification, GEFF was also used for identification of cognitive states, i.e. to identify the task associated to a given FC, regardless of the subject being already in the *Learning* dataset or not (*subject-independent task-identification*). In addition, we also show that eigenvectors from the *Learning Stage* can be characterized as task-dominant, subject-dominant or neither, using two-way ANOVA of their corresponding loadings, providing a deeper insight into the extent of variance in functional connectivity across individuals and cognitive states.

## 1. INTRODUCTION

To date, most studies using fMRI rely on group level analysis where data is averaged over subjects within groups[1–3], potentially ignoring any intra-group individual variability[4]. However, improved acquisition parameters and the increased availability of large datasets[5–11] with open data policy have generated opportunities for the development of subject-level biomarkers from fMRI, thus opening the possibility of personalized medicine for neuro/psychiatric disorders[12]. As clinically useful subject level biomarkers must have high inter-subject differentiabiltiy, also known as subject fingerprint, recent efforts have gone into capturing and improving individual variability in biomarkers based on *functional connectivity* in fMRI data [4,12–14]. Subject- and task-specific signatures have also been found using whole brain *effective connectivity*[15] and *dynamic* functional connectivity[16,17].

Whole-brain functional connectivity patterns are showing increasing promise as subject-level biomarkers that can be estimated from fMRI data. These patterns can be summarized in the form of a full symmetric correlation matrix denominated Functional Connectome (FC). The development of the FC has given birth to the field of *brain functional connectomics* which has been extensively used to study brain connectivity across a wide range of brain disorders[1,18,19]. Recently, it has been shown that FCs have a recurrent and reproducible individual fingerprint [12–14,20], that can be used to identify an individual from a population of FCs. We refer to this process as *subject-identification (SI)*. Using data from the Human Connectome Project (HCP), individual fingerprints have been shown to exist in all eight different tasks (resting-state (RS); emotion (EM); gambling (GAM); language (LAN); motor (MOT); relational (REL); social (SOC) and working memory (WM)), but, apart from resting-state, the SI accuracy was moderate to low[21].

Following the discovery of a fingerprint in FC, Amico and Goñi introduced the "Identifiability Framework ($I_f$)[22] which improved the SI accuracy for all eight tasks from the HCP dataset. Using group-level Principal Component Analysis (PCA) decomposition of FCs, the framework works as a denoising procedure that uncovers latent fingerprints; noisy principal components were identified (and removed) by maximizing *differential identifiability* (similarity of an individual's FC across two sessions, relative to its similarity to the rest of the population). This denoising based on maximizing *differential identifiability* not only improves SI accuracy, but also the capacity to predict fluid intelligence from FCs[22]. This framework has been tested to improve individual fingerprint for different scanning lengths[22], across scanners, with and without global signal regression[23], and across network properties[24]. An extension of this framework has been also used to assess disease progression[25].

Although promising, the existing frameworks[21,22] used for *subject-identification* are not *task-independent*, meaning that an FC from one task cannot be used to identify an individual from a population of FCs from another task even with moderate accuracy rates. Even though the *differential identifiability framework* improves the SI accuracy for each individual task, it does not make the SI process any more *task-independent*. This could be the result of the differential identifiability framework trying to make FCs within tasks as similar as possible, thus potentially removing components which could help with identification across tasks.

In addition to subject fingerprint, functional connectivity patterns, and in turn FCs, have also been shown to vary depending on the cognitive state[26,27] of an individual (i.e. *task-fingerprinting*)[28–32]. Thus, *task-identification* (TI), or the ability to identify the task associated with a given FC from a population of reference FCs that include a collection of tasks, has also become a key goal in the field of brain connectomics. Task identification frameworks have been recently proposed by Xie et al.[16], Pallarés et al.[15] and more recently, Wang et al.[33] using *dynamic* functional connectivity, effective connectivity, and deep learning, respectively. Although useful, these frameworks present some challenges. While effective connectivity showed improved identification performance with respect to functional connectivity, it requires not only functional connectivity but also structural connectivity and a mathematical model of cortical dynamics with its corresponding parameters. Dynamic functional connectivity (dFC) suffers from a subjective and data dependent choice of window length[34]. Deep learning frameworks, although effective in some cases, are *black boxes* and difficult to generalize to new datasets[35]. In contrast, static functional connectivity is easier to compute and is being widely used in the network neuroscience community. Existing TI frameworks are either subject dependent[15] or can only perform task-fingerprinting at the group-level, after removing the subject-specific fingerprints (specific independent components) from the data[16]. Thus, the field still lacks a framework that can perform *task-identification* on functional connectivity while still preserving individual level variability necessary for personalized medicine.

Both subject and task identification can be thought of as object recognition problems. Eigenspace embedding[36] is a common technique used in object recognition, detection, and tracking due to its simplicity and effectiveness. Essentially, high dimensional training images are used to create a low dimensional eigenspace. Then, both training images and target objects are projected into this low dimensional eigenspace and distances are computed between target and training images to detect and/or track certain objects. A number of techniques based on this basic principle have been developed to detect and recognize human faces[37,38], recognize 3D objects and estimate their pose[39], and identify partially occluded objects and estimate their pose[40]. In short, it is a low cost (in terms of memory space and processing time) and computationally efficient image recognition method.

In this study, we propose a framework based on eigenspace embedding for functional connectome fingerprinting (GEFF). Instead of images, whole-brain functional connectomes (FCs) are embedded into a low dimensional eigenspace and classified based on subjects or tasks. Our aim is to achieve four major goals: (i) increase the SI accuracy, (ii) make the SI process potentially task-independent, (iii) perform TI process with high accuracy and, (iv) make the TI process subject-independent, while preserving individual level variability in FCs. In essence, we introduce a fingerprinting framework that, given an FC for a particular individual performing a particular task, is able to identify the subject and/or task with high accuracy.

## 2. METHODS

### 2.1. *Dataset*

The fMRI dataset used in this study is from the publicly available Human Connectome Project (HCP). Per HCP protocol, written informed consent was obtained from all subjects by the HCP Consortium. Full description of the acquisition protocol and processing steps is given below.

### 2.2. *HCP: Functional Data*

We assessed the 100 unrelated subjects (54 females, 46 males, mean age = 29.1 ± 3.7 years) from the HCP 900 subjects data release[6]. This subset of subjects was chosen from the overall dataset to ensure that no two subjects are family relatives. The criterion to exclude family relatives was crucial to avoid confounding effects in our analyses due to family-structure co-variables. The resting-state fMRI scans were acquired on two different days, with two sessions each with two different acquisitions (left to right or LR, and right to left or RL)[5]. The seven fMRI tasks were: emotion, gambling, language, motor, relational, social, and working memory. The gambling, motor and working memory tasks were acquired on the first day, and the emotion, language, relational and social tasks were acquired on the second day. The HCP scanning protocol was approved by the local Institutional Review Board at Washington University in St. Louis. For resting-state fMRI, only the two sessions from day one were used in this study. Full details on the HCP dataset have been published previously[5,41,42].

## 2.3. Brain Atlas

A multi-modal parcellation of the human cerebral cortex, with 180 brain regions in each hemisphere (360 total), was used in this work[43]. For completeness, 14 subcortical regions were added, as provided by the HCP release (filename Atlas_ROI2.nii.gz). To do so, this file was converted from NIFTI to CIFTI format using the HCP workbench software[43,44] (http://www.humanconnecome.org/software/connectome-workbench.html, command -cifti-create-label).

## 2.4. HCP Preprocessing: Functional Data

The data processed using the 'minimal' preprocessing pipeline from the HCP was employed in this work[41]. This pipeline included artifact removal, motion correction, and registration to standard space. Full details on this pipeline can be found in earlier publications[41,42]. The main steps were spatial (minimal) preprocessing, both in volumetric and grayordinate space (i.e. where brain regions are mapped onto the native mesh cortical surface)[42]; slice-timing correction; minimal high-pass temporal filtering (using the -*bptf* option in FSL's[45] *fslmaths* tool; 2000*s* full width at half maximum) applied to both volumetric and grayordinate forms, effectively removing linear trends in the data (no low pass filtering was applied in this 'minimal' HCP pipeline); MELODIC ICA[45] applied to volumetric data; and using FIX[46] to identify and remove artifact components. Artifacts- and motion-related time courses were regressed out (i.e. the six rigid-body parameter time series, their backwards-looking first differences, and the squares of all 12 resulting regressors) of both volumetric and grayordinate data[42].

We added the following steps to the 'minimal' HCP processing pipeline. For resting-state fMRI data: (i) we regressed out the global gray-matter signal from the voxel time courses[47], (ii) we applied a bandpass first-order Butterworth filter in forward and reverse directions (0.001Hz to 0.08Hz[47]; MATLAB functions *butter* and *filtfilt*), and (iii) the voxel time courses were z-scored and then averaged per brain region, excluding any outlier time points that were outside of 3 standard deviation from the mean (*workbench* software, command -*cifti-parcellate*). For task fMRI data, we applied the same steps as mentioned above but a more liberal frequency range was adopted for the band-pass filter (0.001Hz to 0.25Hz), since the connection between different tasks and optimal frequency ranges is still unclear[48].

## 2.5. Estimating Individual Functional Connectomes

Pearson correlation between the time courses of all possible brain region pairs (MATLAB command *corr*) results in a symmetric correlation matrix for each fMRI session of each subject. In this paper we would refer to this object as Functional Connectome (FC). Each task has two sessions — one with left-to-right (LR) and the other with right-to-left (RL) phase-encoding. To avoid any session bias, for each task separately, FCs were chosen randomly from LR and RL sessions such that we had equal number of FCs from both in the two sessions. Finally, the resulting individual FCs were ordered according to the seven so called 'yeo' Functional Networks (FNs), as proposed by Yeo and colleagues[49]. For completeness, an eighth FN comprising the 14 HCP subcortical regions was added (as analogously done in recent papers[22,50]). This reordering was done for visualization purposes only, so that any visualization of FCs or FC-related objects would be somewhat visually interpretable.

## 2.6. Mathematical Notations

In this section, we would establish a few mathematical notations that would be used throughout the paper. Scalar is an italicized letter e.g. $a$. A vector is denoted by a bold italicized letter e.g. $\boldsymbol{a}$, which would be a column vector by default unless otherwise specified. Matrix is denoted by a capitalized italicized bold letter e.g. $\boldsymbol{A}$. For any given vector $\boldsymbol{a}$, the average of its entries is denoted by $\langle \boldsymbol{a} \rangle$, while its norm or magnitude is denoted by $\|\boldsymbol{a}\|$.

If $r \in [q]$, it means that $r$ accepts integer values from 1 up to $q$, where $q \in \{all\ positive\ integers\}$.

Finally, if the $i^{th}$ sample of a set S with cardinality N has a class label where the set of class labels is $[q] = \{1,2, \dots, q\}$, then it would be denoted by $s_i \in [q]^N$.

## 2.7. *GEFF: A framework for Graph Embedding for Functional Fingerprinting*

The GEFF framework consists of two stages: *Learning* and *Identification*. In the *Learning* stage, we compute an eigenspace representation of each learning FC using group-level Principal Component Analysis (PCA)[51,52] decomposition. In the *Identification* stage, we first compute average representations (centroids) of each underlying class in the learning dataset. Then, using the eigenvectors computed in the learning stage, we project each validation FC into the eigenspace and identify it by matching it with one of the class centroids (Figure 1). It has to be noted that GEFF is somewhat similar in its setup with the "Identifiability Framework ($If$)" proposed by Amico and Goñi[22], but there are key differences. First, there is no reconstruction in GEFF and all the processing takes place in the eigenspace. In addition, as opposed to the $If$, GEFF does not require two runs (test/retest FCs) of the same subject in its setup. The two stages of GEFF are described in detail below.

### 2.7.1. Learning Stage: Eigenspace Embedding

An FC is an $m \times m$ symmetric correlation matrix ($m$ is the number of brain regions in the parcellation), and hence can be vectorized into a $M = m(m-1)/2$ dimensional vector by taking the upper triangular part of the matrix (excluding the main diagonal). Analogously to Amico and Goñi[22], we vectorized all the learning FCs and organized them into a matrix

$$X = [x_1, x_2, \ldots, x_N]$$

*where*
$x_i$ *is an M-dimensional vectorized learning FC ($i \in [N]$), and*
*N is the number of learning FCs.*

To construct an eigenspace, we create a PCA decomposition of the input matrix $X$ (MATLAB command *pca*) to extract the *eigenvectors* and the representations (projections) of $x_i$ vectors in(to) the eigenspace.

Analytically, eigenvectors are obtained by solving the following equation:

$$\overline{X}\overline{X}^T u_i = \lambda_i u_i$$

*where*
$\overline{X} = [x_1 - \langle x_1 \rangle, x_2 - \langle x_2 \rangle, \ldots, x_N - \langle x_N \rangle]$, *and*
$u_i$ *represents an M-dimensional eigenvector of the $\overline{X}\overline{X}^T$ covariance matrix, with a corresponding eigenvalue $\lambda_i$.*

Eigenvectors $U = [u_1, \ldots, u_N]$ are arranged in descending order of their eigenvalues, which is equivalent to descending order of their explained variance. For any value of $k \leq N$, the $M$-dimensional vectorized FC $x_i$ can be projected to the eigenspace using the following equation:

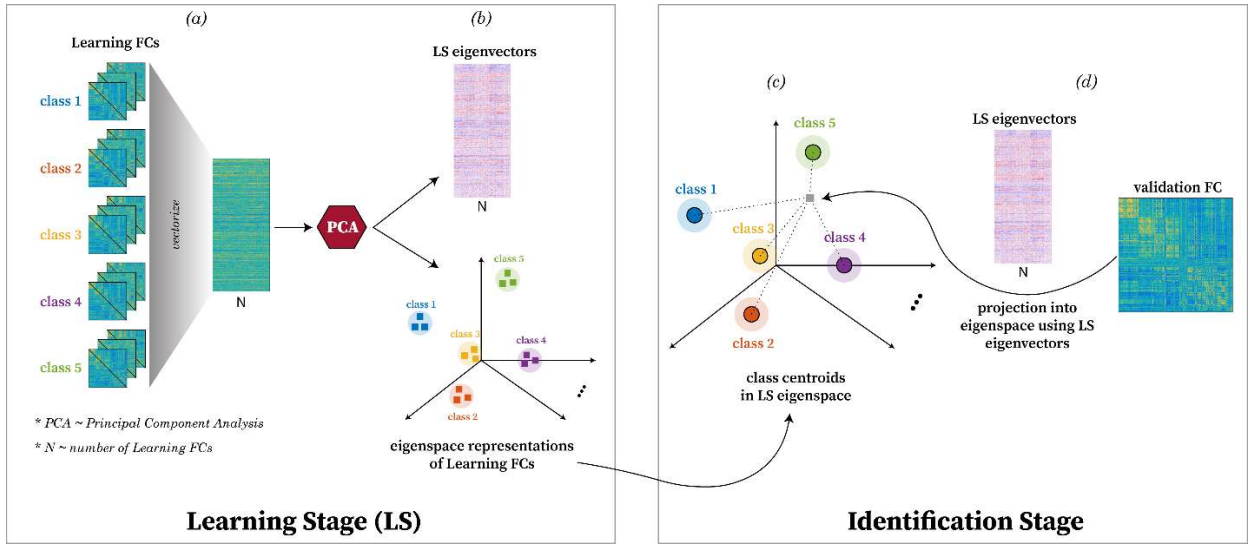

**Figure 1: GEFF, the identification framework.** GEFF consists of two stages: Learning and Identification. During the Learning Stage, all learning FCs are vectorized, organized together (a) and then projected into the eigenspace using PCA (b). During the Identification Stage, we compute average representations (centroids) of each underlying class in the learning dataset (c). Then each validation FC is projected into the eigenspace using eigenvectors from the Learning Stage (d) and is identified by matching its projection with one of the class centroids (c).

$$y_i^k = [u_1, \ldots, u_k]^T \overline{x}_i$$

*where*
$\bar{x}_i = x_i - \langle x_i \rangle$, *and*
$y_i^k$ *is the k-dimensional representation of* $x_i$ *in the eigenspace.*

Using this procedure, we obtained $k$-dimensional representations for all learning FCs, for $k = 1, 2, \ldots, N$.

### 2.7.2. Identification Stage: Nearest Centroid Classifier

The identification process is essentially a multi-class classification problem where the objective is to label an FC in the validation data to one of the classes in the learning data. In this work, we used the Nearest Centroid Classification with the idea that an average representation of a class (subject or task) would be more robust and generalizable than individual samples of that class.

For a given value of $k \in [N]$, we had class-labeled learning samples i.e. $\{(y_1^k, z_1), \ldots, (y_N^k, z_N)\}$,

*where*
$y_i^k$ *is the k-dimensional eigenspace representation of the i-th learning FC (i $\in$ [N]),*
*and* $z_i \in [Z]^N$ *is the corresponding class label.*

Using these samples, we computed per-class centroids:

$$c_l^k = \frac{1}{|C_l|} \sum_{i \in C_l} y_i^k$$

*where*
$c_l^k$ *is the k-dimensional centroid of class* $l \in [Z]$,
$C_l$ *is the set of indices of samples belonging to the class* $l \in [Z]$, *and*
$|C_l|$ *is the number of samples or size of the class* $l \in [Z]$.

For SI and TI processes, classes correspond to the subjects and the tasks included in the learning dataset, and these centroids are *average* representations of the subjects and tasks in the eigenspace, respectively.

For a given validation FC, we first vectorized it into an $M$-dimensional vector $w$. We then obtained a $k$-dimensional vector $g^k$ by projecting $w$ to the eigenspace constructed in the learning stage using the following equation:

$$g^k = [u_1, \ldots, u_k]^T \bar{w}$$

*where* $\bar{w} = w - \langle w \rangle$.

To provide an alternative and perhaps more intuitive perspective, one may also think of this process as a multi-linear regression:

$$\bar{w} = U_k \beta + \varepsilon$$

*where*

$\overline{w}$ is the dependent variable or the validation FC,
$U_k = [u_1, ..., u_k]$ represents the transposed independent variables or the eigenvectors,
$\beta = g^k$ is the k-dimensional vector of estimated coefficients, and

$\varepsilon$ is the residual noise.

A validation FC ($g^k$) was identified as belonging to class $l^*$ that minimized the distance between $g^k$ and the class centroid $c_l^k$:

$$l^* = \underset{l}{argmin}\, dist(g^k, c_l^k)$$

where
$c_l^k$ is the centroid for class $l \in [Z]$, and
'dist' represents the distance function that was used to compute distance between the input $g^k$ and the class centroids.

In our case, we used the *cosine* distance which is given by the following equation:

$$dist(x, y) = 1 - cos\theta = 1 - \frac{\langle x, y \rangle}{\|x\|\|y\|}$$

where
$\langle x, y \rangle$ is the dot product of vectors $x$ and $y$.

For high dimensional data, to measure closeness or distance between two unit vectors, a natural choice, empirically, would be the angle between them, or the *cosine* of that angle[53]. Although the framework was also tested with correlation distance and Euclidean distance and similar results were found (results not shown).

We repeated the identification process for all the validation FCs and the identification rate was defined as

$$Identification\ Rate = \frac{Number\ of\ correctly\ labeled\ validation\ FCs}{Total\ number\ of\ validation\ FCs}$$

Using this generic definition of accuracy, we can also compute SI or TI rates for subsets of validation dataset for all possible values of eigenspace dimensionality $k = 1, 2, ..., N$, which depends on how FCs are split into learning and validation data and is described in more detail below.

### 2.8. Subject Identification (SI) Process

For each of the 100 unrelated subjects, we had eight different fMRI tasks (including resting-state), as described above. For each task, we had two runs, here referred to as Test and Retest. For resting-state, we had four runs in total, two runs per session, but we only used the two runs from session 1 to balance the dataset with task fMRIs. For simplicity, we will refer to resting-state as a task, unless stated clearly otherwise.

For SI statistics, we must consider the dependence between subjects in the sample. For instance, if two subjects A and B are very close to each other, B might be misclassified as A. But, if A was not in the learning dataset, it is possible that B would have been classified correctly. A convenient procedure to assess variability in the identification process is to use random cross-validation resampling[54], with each resample comprising random draws without replacement of the box containing the group of subjects.

Within every cross-validation run, we randomly picked 80% of the 100 subjects ($100 * 0.8 = 80$) as our learning subjects from the *Test* session (*It must be emphasized that for any task, FCs from the two runs (left-to-right vs right-to-left phase or LR vs RL) of the original HCP dataset were randomly assigned to either the Test or the Retest. That is why choosing FCs from only Test is essentially choosing FCs randomly from the two available sessions of LR and RL*). For every subject, we picked $T \in [7]$ number of task FCs, which resulted in $N = 80 * T$ FCs in the learning dataset i.e.

$$X = [x_1^1, x_2^1, \ldots, x_T^1, \ldots \ldots \ldots, x_1^{80}, x_2^{80}, \ldots, x_T^{80}]$$

where
$x_i^j$ is the vectorized FC for the jth subject and the ith task.

Then, as described above, we apply PCA to $X$ in order to create an eigenspace and compute $k$-dimensional eigenspace representations for all the learning FCs for a given value of $k$ ($k$ represents the eigenspace dimensionality or the number of eigenvectors chosen for the projection in the order of descending eigenvalues or equivalently, explained variance) i.e.

$$Y^k = [y_1^1, y_2^1, \ldots, y_T^1, \ldots \ldots \ldots, y_1^{80}, y_2^{80}, \ldots, y_T^{80}]$$

where
$Y^k$ is the matrix of k-dimensional projections of all the learning FCs.

For the SI process, 'subjects' are the classes i.e. $Z = 80$. So, in the Identification Stage, one centroid is computed per subject i.e.

$$C_{subj}^k = \{c_1^k, c_2^k, \ldots, c_{80}^k\}$$

where
$C_{subj}^k$ is the matrix of all subject-centroids in the k-dimensional eigenspace.

These centroids reflect an average representation of subjects across tasks which was then utilized in the identification process.

In each cross-validation resample, the *validation* dataset comprised of new FCs (additional runs of the learning tasks or external tasks) of the same subjects employed in the learning dataset. FCs in the validation dataset always included all tasks for all learning subjects. Hence, overall it always comprised of $80 * 8 = 640$ FCs. The validation dataset was subdivided into two categories:

**1) Within-Learning-Tasks:** new FCs that belonged to the tasks that were included in the learning dataset

**2) Across-Tasks:** new FCs that belonged to the tasks that were *not* included in the learning dataset.

All the validation FCs were projected into the eigenspace and were labelled by identifying the nearest 'subject centroid' as described in detail in the Methods section.

The SI process was performed for:

1) Within-Learning-Tasks and Across-Tasks, separately
2) 100 random cross-validation resamples
3) all the values of eigenspace dimensionality i.e. $k = 1, 2, ..., N$, and
4) different number of learning tasks i.e. $T = 1, 2, ..., 7$. For a given value of $T$, the process was repeated for all possible permutations of tasks in the learning dataset. For instance, if $T = 2$, there are $\binom{8}{2} = 28$ possible permutations in which we can pick two tasks out of eight. So, the process was repeated for all 28 permutations.

## 2.9. Task Identification (TI) Process

Like the SI process, we must consider the dependence between subjects in the sample. Although here the consideration is slightly different. Two subjects A and B from the same task when averaged, could create a 'better' average representation of the task than say subjects B and C. Here the word 'better' means a representation that is more generalizable to the rest of the sample and hence would perform better in the identification stage. As done during the SI process, variability in the identification process was assessed by using random cross-validation resampling[54], with each resample comprising random draws of subjects without replacement.

Additionally, we should consider the number of subjects per task in the learning dataset, because intuitively a larger sample of subjects per task could create a 'better' average representation of the task than a smaller one. So, we need to explore the TI process over a range with respect to the number of subjects per task in the learning dataset.

Within each cross-validation run, $n$ number of subjects from the *Test* session are chosen randomly per task. So, the total number of FCs in the learning dataset would be $N = 8n$, since there are in total 8 tasks i.e.

$$X = [x_1^1, x_2^1, ..., x_n^1, ......... , x_1^8, x_2^8, ..., x_n^8]$$

*where*
$x_i^j$ *is the vectorized FC for the jth task and the ith subject.*

Then, just as we did in the SI process, an eigenspace was created using PCA and all the learning FCs were projected into the eigenspace for a given value of $k$ i.e.

$$Y^k = [y_1^1, y_2^1, ..., y_n^1, ... ... ..., y_1^8, y_2^8, ..., y_n^8]$$

where
$Y^k$ is the set of k-dimensional projections of all the learning FCs.

For the TI process, classes are the different 'tasks', instead of 'subjects' i.e. $Z = 8$. So, in the Identification Stage, one centroid was computed per task i.e.

$$C_{task}^k = \{c_1^k, c_2^k, ..., c_8^k\}$$

where
$C_{task}^k$ is the set of all task-centroids in the k-dimensional eigenspace.

These centroids reflect an average representation of tasks across subjects which was then utilized in the identification process.

The *validation* dataset comprised of the FCs from the *Retest* session for all the subjects and all the tasks ($100 * 8 = 800$). The validation dataset was subdivided into two categories:

**1) Within-Learning-Subjects:** new FCs that belonged to the same subjects that were included in the learning dataset, and

**2) Different-Subjects:** new FCs that belonged to all the other subjects that were *not* included in the learning dataset.

All the validation FCs were projected into the eigenspace and were labelled by identifying the nearest 'task' centroid.

The SI process was performed for:

1. Within-Learning-Subjects and Different-Subjects
2. 100 cross-validation resamples
3. all the values of eigenspace dimensionality i.e. $k = 1, 2, ..., N$, and
4. different number of subjects per task i.e. $n = [2:1:20, 30:10:80]$.

*Null Model Evaluation for the framework*

For both the SI and the TI processes, a null model was evaluated by randomly permuting the class labels of the learning dataset and repeating the identification process.

2.10. *Comparative Analysis: SI and TI using original FCs*

As a comparative analysis, the SI and TI processes were also performed using the original FCs (Orig FCs). The *learning* and *validation* datasets were created the same way and the process was repeated for the same values of different parameters. Instead of averaging the eigenspace representations, Orig FCs were averaged across tasks and subjects for the SI and the TI processes, respectively. The second major difference was in the way the FCs in the validation dataset were compared to the learning dataset.

First, the averaged representations of subjects or tasks (for SI and TI respectively) were vectorized and organized into a matrix i.e.

$$C_{subj} = [c_1, c_2, ..., c_{80}] \text{ (for SI)}$$

and $C_{task} = [c_1, c_2, ..., c_8]$ (for TI)

*where $c_i$ is an averaged FC.*

All the FCs in the validation dataset were also vectorized. A given vectorized validation FC, $y$, was identified as belonging to class $l^*$ that maximized the similarity between the input $y$ and the averaged FC for the class centroid $c_l$:

$$l^* = \underset{l}{argmax}\, d(y, c_l)$$

*where*
*$l \in [Z]$, the set of all class labels,*
*$c_l$ is the averaged FC for the l-th class, and*
*$d(y, c) = \frac{\Sigma_j(y_j - \bar{y})(c_j - \bar{c})}{\sqrt{\Sigma_j(y_j - \bar{y})^2}\sqrt{\Sigma_j(c_j - \bar{c})^2}}$ is the Pearson's correlation coefficient[55,56] between $y$ and $c$.*

A direct comparison between traditional process of identification (for instance Finn et al.[21] or Venkatesh et al.[20]) with GEFF is only possible when we use only one FC per subject in the learning stage. For two or more FCs per subject in the learning dataset, we used averaged-across-tasks FCs instead, as described in detail above. This was necessary so we could keep the comparative analysis with Orig FCs consistent with GEFF, but also make qualitative comparisons with the previous literature.

## 2.11. Characterization of Eigenvectors in Terms of Their Subject- and Task-fingerprint

We did a post-hoc analysis to characterize each eigenvector separately in terms of its subject- and/or task fingerprint. The idea was to see if eigenvectors, separately, indeed hold subject- and/or task-fingerprint and if there are different regimes of eigenvectors based on subject- and task-specificity.

For this process, FCs for all the subjects and for all the tasks from the *Test* and the *Retest* session ($1600 = 2 * 100 * 8$ FCs) were vectorized and then organized into a matrix $X$:

$$X = [x_1, x_2, ..., x_N]$$

*where*
*$x_i$ is an M-dimensional vectorized FC ($i \in [N]$), and*

$N = 2*100*8 = 1600$ is the total number of FCs.

To construct an eigenspace, we input $X$ to PCA (MATLAB command *pca*) to extract the *eigenvectors* and the representations (projections) of $x_i$ vectors in(to) the eigenspace (much in the same way as we did in the Learning Stage for GEFF, Figure a-b):

$$U = [u_1, u_2, ..., u_N], \text{ and}$$

$$Y = [y_1^N, y_2^N, ..., y_N^N]$$

where
$u_i$ is an M-dimensional eigenvector,
$y_i^N$ is the N-dimensional projection of the M-dimensional vector $x_i$ into the N-dimensional eigenspace, and ($i \in [N]$).
The matrix $Y$ can be expanded as:

$$Y = \begin{bmatrix} y_{11}^N & y_{21}^N & \cdots & y_{N1}^N \\ y_{12}^N & y_{22}^N & \cdots & y_{N2}^N \\ \vdots & \vdots & \ddots & \vdots \\ y_{1N}^N & y_{2N}^N & \cdots & y_{NN}^N \end{bmatrix}$$

where each column is an N-dimensional projection ($y_i^N$) in the N-dimensional eigenspace.

These projections can be thought of as coordinates in an *N*-dimensional eigenspace, spanned by the *N* eigenvectors. Hence, the *i*-th row contains the weights or loadings of all the projections corresponding to the *i*-th eigenvector. Since each column corresponds to an FC that belongs to a specific task or a subject, the weights corresponding to each eigenvector can also be grouped by tasks or subjects.

We characterized each eigenvector individually, in terms of its subject- and/or task-fingerprint, using two-way ANOVA on the corresponding weights, where the group effects were 'task' and 'subject'. This analysis was repeated for all 1600 eigenvectors and the corresponding *p*-values and effect sizes (*F*-stats) were computed. The *p*-values were corrected for multiple comparisons using *Bonferroni* correction across the 1600 ANOVAs performed. An eigenvector was declared task- and/or subject-dominant if the corresponding *p*-values was $0.01 (Bonferroni\ corrected)$ and subsequently based on the magnitude of corresponding *F*-stat.

## 3. RESULTS

In this study, we proposed the Graph Embedding for Functional Fingerprinting (GEFF) framework. GEFF was employed to perform *subject-* and *task-identification* (SI and TI, respectively) using the 100 unrelated subjects from the HCP 900 subject data release. GEFF consisted of two stages: 1) Learning and 2) Identification. In the *Learning stage*, we computed an eigenspace representation of each FC in the learning dataset using group-level PCA decomposition. In the *Identification stage*, we computed average representations (centroids) of each underlying class (subjects or tasks) in the learning dataset. Then, using eigenvectors computed in the Learning Stage, we projected each validation FC into the eigenspace and identified it by matching its projection with one of the class centroids (Figure 1).

Both the SI and TI processes were repeated using original FCs (Orig FCs), where average representations of the underlying classes (subjects or tasks) were computed by averaging the corresponding FCs. The class of each validation FC was identified by matching it (using correlation; see **Section 2.10.** for details) with one of these averaged FCs.

### 3.1. Subject Identification (SI)

SI process was performed using different number of task FCs per subject in the learning dataset, which we labeled as $LS^{(i)}$, $i = 1,2,...,7$. To assess the robustness of the results and statistical comparisons between the two frameworks (Orig FCs and GEFF), SI rates were computed for 100 random cross-validation resamples. For each cross-validation resample, 80% of the subjects (for each learning task) were randomly chosen without replacement from the Test session to create the learning dataset. SI rates were then computed for new FCs of the same subjects when 1) FCs belonged to same tasks as the learning tasks (Within-Learning-Tasks) and 2) when FCs belonged to tasks different than the learning tasks (Across-Tasks).

Whenever possible, we show the SI rates separately for the cases where resting-state was part of the learning dataset (RS+) from the cases where it was not (RS−). Even though this choice is somewhat intuitive considering resting-state fMRI is by design different than task fMRIs, we will provide a more practical reason when we discuss the SI results with *two* task FCs per subject in the learning dataset i.e. $LS^{(2)}$.

We should also highlight that variation around the mean behavior (whether across cross-validation resamples or learning tasks permutations) was so small (in most cases) that it was hidden behind the mean solid lines.

*3.1.1. SI using only one Learning Task: $LS^{(1)}$*

At the maximum eigenspace dimensionality, GEFF improved SI rates over Orig FCs for each task and for both Within-Learning-Task and Across-Tasks scenarios (Figure 2a, 2d). Within-Learning-Task SI rates were 90% for GEFF using resting-state, gambling, language, relational, and social tasks. For resting-state, SI rate was exactly 100% across all the cross-validation resamples (Figure 2a). Even for emotion, motor, and working memory task, where the SI rates were lower than 90%, they were still significantly higher than their Orig FCs counterparts (e.g. an improvement of around 30% for motor task) (Figure 2a). Even though Across-Tasks SI rates were considerably lower (with the highest for relational and working memory tasks: ~60%), they were significantly higher than SI rates using Orig FCs (Figure 2d).

SI rates increased monotonically with increasing eigenspace dimensionality (Figure 2b-c, 2e-f). Interestingly, Within-Learning-Task SI rates for resting-state saturate at 100% using only 75% (60/80) of maximum eigenspace dimensionality (Figure 2b). For Within-Learning-Task SI rates, GEFF required less than half of the maximum eigenspace dimensionality to cross the Orig FCs SI rates (Figure 2b-c). On the other hand, GEFF Across-Tasks SI rates required more than half but less than 75% of the maximum eigenspace dimensionality to cross the Orig FCs SI rates (Figure 2e-f).

### 3.1.2. SI using two Learning Tasks: $LS^{(2)}$

At the maximum eigenspace dimensionality, Within-Learning-Tasks SI rates for GEFF were 98% for all permutations across learning tasks (Figure 3a). Interestingly, SI rates for validation FCs from resting-state were considerably lower when resting-state was *not* part of the learning dataset (Figure 3b). On the other hand, if we included resting-state in the learning dataset, along with one other task, we saw that SI rates for all the validation tasks were very high, whether those tasks were part of the learning dataset (Figure 3a) or not (Figure 3b). Combination of resting-state with motor task in the learning dataset seemed to be an exception as it resulted in lower SI rates for relational and social task (70 − 80%). This special behavior of resting-state compelled us to separate the cases where resting-state was part of the learning dataset from cases where it was not.

Just as observed with one learning task, SI rates increased monotonically with increasing dimensionality (Figure 3c-f). Within-Learning-Tasks SI rates saturated at 98% using only 75% (120/160) of maximum eigenspace dimensionality when resting-state was *included* in the learning dataset (RS+; Figure 3c), and at 92% when resting-state was *not* included (RS−; Figure 3d). When resting state was *not* included in the dataset, average Across-Tasks SI rates were 80% (RS−; Figure 3f) and increased to 90% when resting-state *was* included (RS+; Figure 3d).

It should be noted that Across-Tasks SI rates never reached a saturation point (Figure 3d, 3f). Also, for both Within-Learning-Tasks and Across-Tasks, GEFF required less than half of the maximum eigenspace dimensionality to cross the Orig FC SI rates (Figure 3c-f). We should also highlight that without RS in the learning dataset, six or more tasks are required to reach similar Across-Task SI rates as with RS and one other task in the learning dataset (Figure S1; bottom row). We explore this in more detail in the next subsection.

At this point, we have shown that using GEFF improved SI rates for all tasks individually ($LS^{(1)}$; Figure 2) and we achieved close to perfect SI rates using only two tasks in the learning dataset ($LS^{(2)}$) when the learning and validation FCs come from the same tasks (Figure 3a). In addition, SI process can be made potentially *task-independent* using only two tasks in the learning dataset, if one of the tasks is resting-state, although the corresponding rates are 90% which can still be improved (Figure 3b, 3d). For this purpose, we considered Across-Tasks SI rates using more than two tasks in the learning dataset.

### 3.1.3. Across-Tasks SI rates using more than two Learning Tasks: $LS^{(i)}$ ($3 \leq i \leq 7$)

With resting-state included in the learning dataset (RS+), we reached Across-Tasks SI rates of 95% with three and 98% with four learning tasks. Beyond that, the improvement in SI rates was marginal (Figure S1; top row). Interestingly, when resting-state was not included (RS−), SI rates do increase with increasing learning tasks, but achieve a maximum of 92% (Figure 4; bottom row) (compared to 98% with only four learning tasks in RS+). With increasing number of tasks in the learning dataset, the percentage of maximum eigenspace dimensionality required to cross the Orig FC SI rates and to achieve saturation, decreased (Figure S1). Finally, Across-Tasks SI rates increased for Orig FC with increasing learning tasks (just like

GEFF) when resting-state is included in the learning tasks (RS+) (Figure S1; top row) but decreased when resting-state is *not* included (RS−) (Figure 4; bottom row).

## Subject Identification using one task in the Learning Stage: LS$^{(1)}$

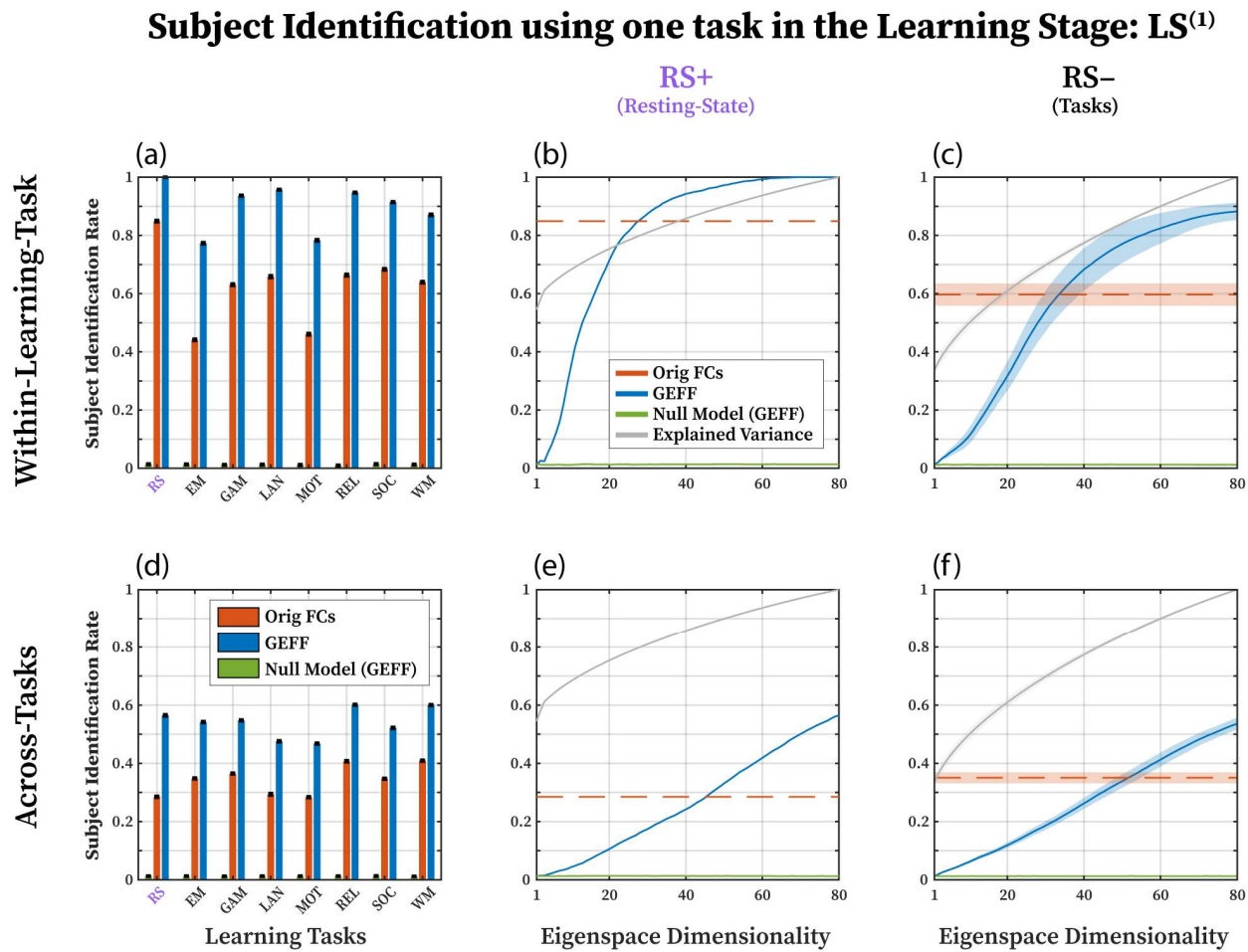

**Figure 2:** Subject-Identification (SI) rates with only *one* task in the Learning dataset (LS$^{(1)}$) for Orig FCs and GEFF. SI rates for each Learning task at the maximum eigenspace dimensionality (i.e. 80) when the *validation* dataset contains new FCs from the same task as the learning stage dataset i.e. *Within-Learning-Task* (a) and when the *validation* dataset is made up of new FCs from tasks *not* included in the learning stage dataset i.e. *Across-Tasks* (d). The bars show the mean and the error bars show the standard error of mean (SEM) across the cross-validation resamples. (b) and (e) show the SI rate curves with increasing eigenspace dimensionality for *Within-Learning-Tasks and Across-Tasks* when only FCs from resting-state are chosen as the learning dataset (RS+). On the other hand, (c) and (f) show similar curves for *Within-Learning-Tasks and Across-Tasks* when resting-state is *not* included in the learning dataset (RS−). The solid lines show the mean SI rate across learning tasks and the shaded regions show the SEM.

### 3.1.4. Summarizing the SI results

GEFF improved the subject identification rates over Orig FCs across the board: 1) whether the validation FCs belong to the same tasks as the learning tasks or not (Within-Learning-Tasks or Across-Tasks) and 2) whether the learning tasks include resting-state or not (RS+ or RS−) (Figure 4). We also show that a qualitatively optimal point for GEFF with respect to subject identification accuracy would be when we have two learning tasks and one of those is resting-state (white asterisk, Figure 4). In addition, we show that an average individual representation, whether it was created using Orig FCs or with GEFF, resulted in a much better individual fingerprint (Figure 4; *Within-Learning-Tasks*) and became more generalizable to external tasks (Figure 4; *Across-Tasks*). Finally, the SI rates for the null model under any condition are very low and almost identical to the chance level of identification (Figure 2-3).

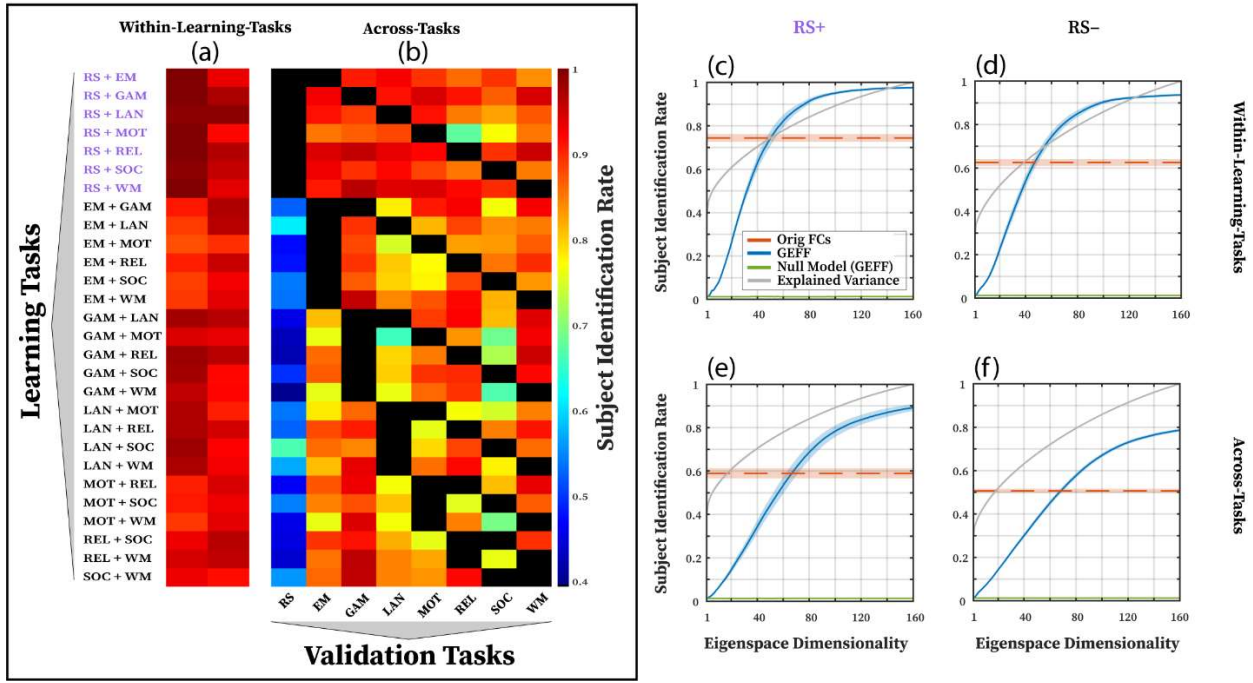

**Figure 3:** *Subject Identification* **(SI) rates with *two* tasks in the Learning stage dataset (LS$^{(2)}$) for Orig FCs and GEFF.** SI rates for each permutation of two tasks in the Learning dataset at the maximum eigenspace dimensionality (i.e. 160) when the *validation* dataset contains new FCs from the same two tasks as the Learning dataset i.e. *Within-Learning-Tasks* (a) and when the *validation* dataset is made up of new FCs from tasks *not* included in the Learning dataset i.e. *Across-Tasks* (b). SI rate curves with increasing eigenspace dimensionality, when one of two tasks in the Learning dataset is resting-state (RS+) are shown in (c) and (e), respectively, for *Within-Learning-Tasks* and *Across-Tasks*. On the other hand, (d) and (f) show similar curves for *Within-Learning-Tasks* and *Across-Tasks* when resting-state is *not* included (RS−) in the Learning dataset. The solid lines show the mean SI rate across all Learning tasks permutations and the shaded regions show the standard error of mean (SEM). Two black rectangles in each row of (b) correspond to the two tasks that were used in the Learning stage for that particular case.

### 3.2. Task Identification (TI)

*3.2.1 TI rate profiles with respect to number of subjects per task*

The first step in TI process was to see how the TI rates change with number of subjects per task in the learning dataset. This process was repeated for a wide range ($n = [2:1:20, 30:10:80]$) of number of subjects per task. To assess the robustness of the results and for statistical comparisons between the two frameworks (Orig FCs and GEFF), TI rates were computed for 100 cross-validation resamples. Within each cross-validation resample, $n$ (where $n = [2:1:20, 30:10:80]$) number of subjects (for all tasks and resting-state) were chosen at random from the Test session to create the learning dataset. TI rates were then computed for new FCs from the same tasks when 1)

FCs belonged to the same subjects as the ones included in the learning dataset (Within-Learning-Subjects) and 2) when FCs belong to all the other subjects that were *not* included in the learning dataset.

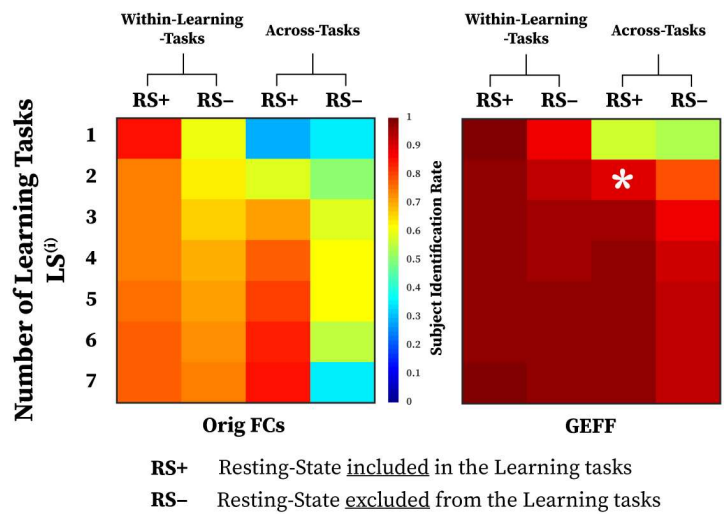

**Figure 4: A summary of Subject Identification (SI) results for Orig FCs (left) and GEFF (right).** For GEFF, the SI rates correspond to the maximum eigenspace dimensionality for a given number of tasks in the learning dataset ($LS^{(i)}$). White asterisk marks a qualitatively optimal setting for GEFF, where both Within-Learning-Tasks and Across-Tasks SI rates are very high while minimizing the learning tasks to 2.

We observed that at 20 subjects per task in the learning dataset, the TI rates reach a plateau for both Within-Learning-Subjects and Different-Subjects, although there was marginal increase with GEFF with increasing number of subjects per task (Figure 5). TI rates using Orig FCs saturated around 91% and were always lower than corresponding TI rates for GEFF which saturated around 99%.

It should be highlighted that TI rates for GEFF were computed at the maximum eigenspace dimensionality for each value of $n$. Also, the standard error of mean across cross-validation resamples was so low that it is hidden behind the mean lines (Figure 5).

*3.2.2. TI using 20 subjects per task*

After establishing that TI rates reach a saturation point after 20 subjects per task, we assessed the TI rates at $n = 20$ in more detail. We observed that TI rates for GEFF cross the Orig FC TI rates with just 27.5% (44/160) and 30% (48/160) of the maximum eigenspace dimensionality for Within-Learning-Subjects (Figure S2) and for Different-Subjects (Figure S3), respectively. We noticed that the TI rates for GEFF saturated after 50% (80/160) of the maximum eigenspace dimensionality at 95% and 94% for Within-Learning-Subjects and Different-Subjects respectively (Figure S2 and S3). Another important observation was that the TI rate rises sharply with the first three eigenvectors and then steadily increases with increasing dimensionality (Figure S2 and S3). This observation highlights the importance of the first few eigenvectors in the TI process, which will be discussed again in the next section (*Characterization of Eigenvectors*). Finally, the confusion matrices shown in Figures S2 and S3 highlight that when the TI rates improve with increasing eigenspace dimensionality, they do so for all the tasks. This also shows that certain tasks (e.g. emotion, gambling, relational) are harder to identify than others (e.g. resting-state, social).

We should also highlight that the TI rates for the null model are very low and almost equal to the chance levels of identification rates (Figure 5; Figure S2-S3).

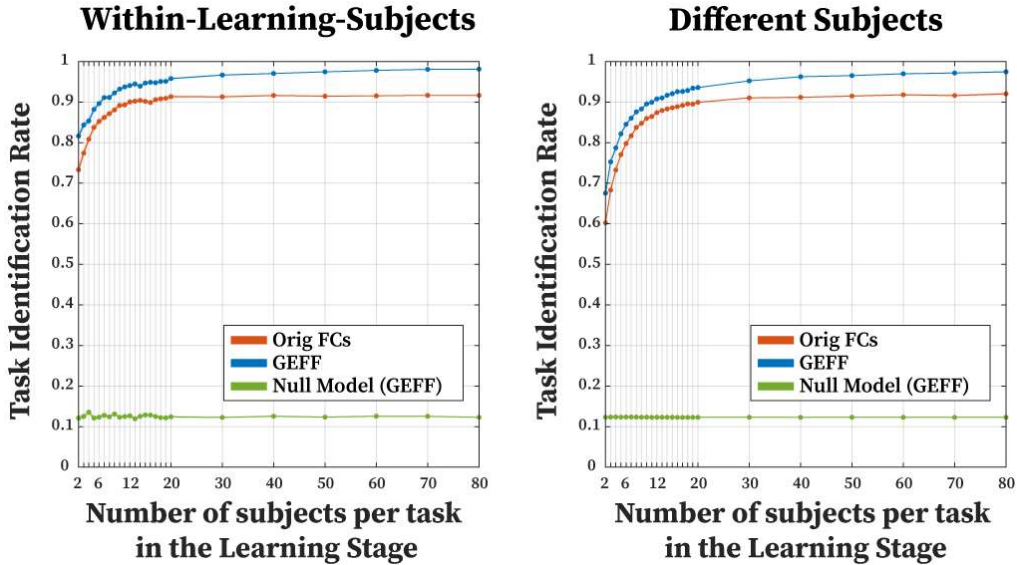

**Figure 5:** **Task Identification (TI) rate curves with increasing number of subjects per task in the Learning Stage dataset for Orig FCs and GEFF.** TI rates shown were computed at the maximum eigenspace dimensionality. Left panel shows the TI rate curves when *validation* dataset contains new FCs from the same subjects as the ones included in the Learning dataset i.e. *Within-Learning-Subjects*. Right panel, on the other hand, shows the TI rate curves when *validation* dataset is made up of new FCs from subjects not included in the Learning dataset i.e. *Different Subjects*. Solid lines with dots show the mean TI rates across cross-validation resamples, while the shaded areas around the mean show the standard error of the mean (SEM). Note that the SEM is so small that it's hidden behind the solid mean lines.

### 3.3. Characterization of Eigenvectors in Terms of Their Subject- and Task-fingerprint

Using all 1600 ($2 * 100 * 8$) FCs from *Test* and *Retest* sessions, we computed the 1600 eigenvectors and their corresponding weights using group-level PCA (see Methods). To ascertain the task- and subject-specificity of a given eigenvector, *two-way* ANOVA was applied to its corresponding weights using 'task' and 'subject' as the two group effects. This process was repeated for all 1600 eigenvectors and the corresponding *p*-values were *Bonferroni* corrected.

We observed that the eigenvectors can be divided into three regimes: 1) Task-Dominant, 2) Subject-Dominant, and 3) Neither (Figure 6). The Task-Dominant regime consists of the first $10 - 20$ eigenvectors which explain $80 - 90\%$ of the variance in the data. Then, we observed a second wave of eigenvectors which constitute the Subject-Dominant regime. This regime lasts till around 300 eigenvectors and is followed by the *Neither regime* which is neither task- nor subject-specific.

It should be noted here that there are no hard boundaries between these regimes. A task dominant eigenvector can have subject-specificity (e.g. the first 10 eigenvectors) and vice versa. However, it is noteworthy that ordering the eigenvectors by their explained variance separated them into

task- and subject-dominant regimes, instead of task- and subject-specificity spuriously distributed across the range of 1600 eigenvectors.

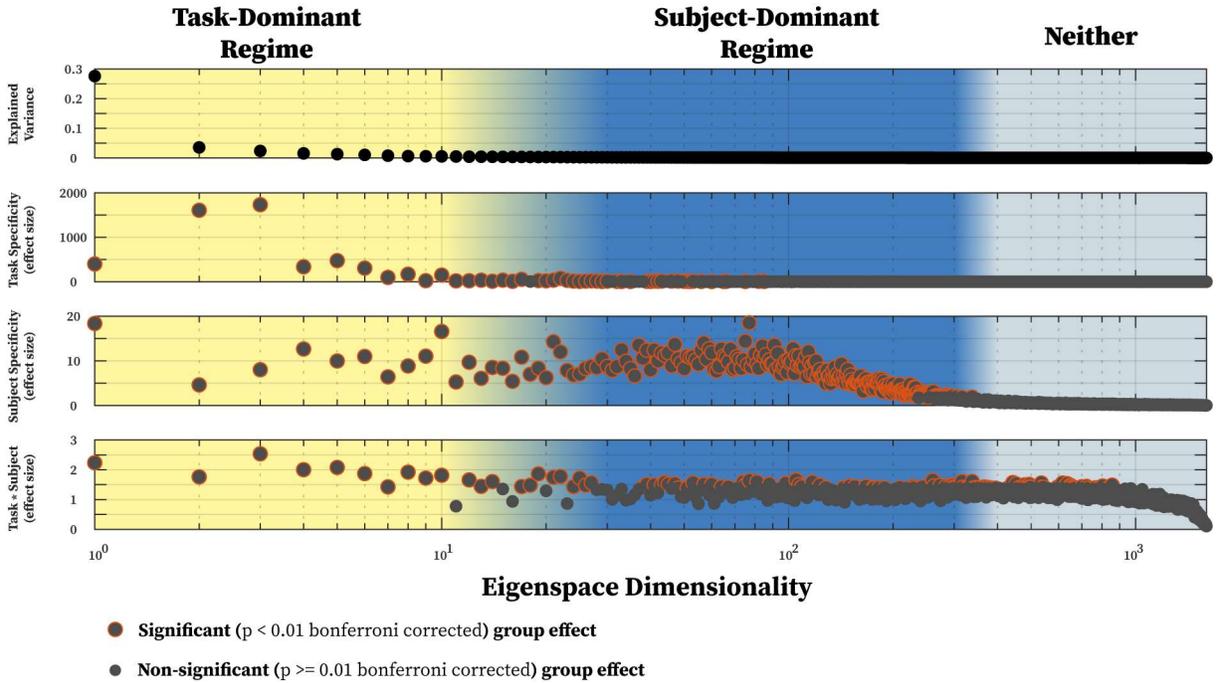

**Figure 6: Characterization of individual eigenvectors.** Top panel with black dots shows the explained variance of each eigenvector individually. The middle and lower panels show the group effects (*F*-statistic) for *task* and *subject* groups, computed using two-way ANOVA on each eigenvector weights. The black dots with orange boundary highlight eigenvectors with significant group effects ($p < 0.01$; bonferroni corrected across the 1600 eigenvectors), while the gray dots show the non-significant ones.

## 4. DISCUSSION

In this paper, we proposed an embedding framework for FC fingerprinting called GEFF: Graph Embedding for Functional Fingerprinting. We employed this framework to perform Subject- and Task-Identification (SI and TI respectively) using functional connectomes. Compared with existing frameworks, not only did GEFF considerably improve the SI and TI rates, it also made the SI and TI processes, respectively, *task-* and *subject-independent*. GEFF proved to be a highly accurate and potentially universal FC fingerprinting framework which allowed us to robustly estimate individual fingerprint and decode cognitive states from FCs. We also showed that resting-state combined with one other task covers the entire cognitive space in terms of individual fingerprinting. We also characterized the learning stage eigenvectors, and found that they can be delineated into task- and subject-dominant regimes by simply arranging them in the descending order of their explained variance.

### *4.1. Creating an average individual representation using multiple task FCs improves individual fingerprint*

An average individual representation, whether it was created using original FCs in the connectivity domain or with GEFF in the eigenspace, resulted in a much better individual fingerprint, especially when the FCs being identified belonged to the same tasks that were used to create the average representations. With only one exception, by adding more tasks to create the average representations, the individual fingerprint became more accurate and generalizable to external tasks. This result aligns with previous work of Gao et al.[57] where they show that combining multiple FCs improves predictive estimates of phenotypic measures.

For original FCs (Orig FCs), there was one exception to this trend. That happened when resting-state was not included to create the average representations and the FCs being identified belonged to the tasks different from the ones used to create the average. The fingerprint became worse when more and more FCs were used to create the average. This is partially explained by the fact that some of the validation FCs that we were trying to identify in these cases were resting-state FCs. As we would explore in more detail later, it is hard to identify resting-state FCs when the average representations are created using only non-resting-state tasks. The more tasks we used to create the average representations, the fewer tasks (including resting-state) were left for identification. This resulted into a higher percentage of resting-state FCs in the validation data, which in turn caused a decrease in the fingerprinting accuracy. We did a post-hoc analysis to investigate this further. When resting-state is removed from the validation FCs, the fingerprinting accuracy increases with increasing number of tasks participating in the average representations (Figure S4). However, this argument didn't hold when 5 or more tasks were used for average representations, as the identification rates slightly decreased for Across-Tasks. In other words, the individual fingerprint became less generalizable to external tasks with 5 or more tasks in the Learning dataset. We should also emphasize that this behavior was only observed for original FCs and when resting-state was excluded while computing the average representations for individuals. With GEFF, the individual fingerprint became always more accurate and generalizable to external tasks when more tasks (with and without resting-state) were used for average individual representations.

### *4.2. Individual fingerprinting is almost perfect with GEFF when individual FCs being compared are from the same tasks*

When performing individual fingerprinting for Within-Task FCs, GEFF exhibited near perfect performance. Using GEFF, the FC fingerprint was universally improved (90% accuracy for most tasks), with a perfect fingerprint (100% accuracy) for resting-state. In addition, when more than one task was used in GEFF to create an average representation of individuals, individual fingerprinting was nearly perfect for any combination of tasks (98% accuracy) as long as the new FCs that were being compared with the average representations belonged to the same tasks as the ones used to create the average representation. This widely outperformed the canonical method for performing FC fingerprinting using correlation between FCs belonging to the same task. Using this canonical approach, only resting-state FCs had a reasonable fingerprint (85% accuracy) while all the other tasks performed poorly (70% for all tasks; emotion and motor 50%). Although canonical fingerprints did improve when we created the average individual representations with original FCs (except for Across-Tasks), GEFF always outperformed by a margin $(10-20\%)$ for all possible number and combination of tasks.

Finn et al.[21] reported a mean identification accuracy of 93.65% for resting-state using Pearson correlation, which is higher than 85% that we obtained (Figure 2a; RS). Recently, Venkatesh et al[20] obtained 77.5% identification rate with RS when using correlation as a similarity metric. Given that the HCP data has four runs for RS, two on each day, Finn et al. averaged the two FCs from the same day into a single FC. Altogether this suggests that averaging across several runs of the same task produces a more representative FC, which results in higher fingerprinting accuracies. In this work, however, we only used the two runs from one of the two days and kept the two runs separate. The identification accuracy still increased to a perfect 100% with GEFF.

Finally, a geometry-aware approach for comparing FCs, within a single task, was recently proposed. This method outperformed the canonical methods of using correlation-based similarity metric across all tasks[20]. It is noteworthy that GEFF outperforms this approach as well across all tasks (e.g. an improvement of around 20% for emotion, gambling, and relational tasks for Subject Identification).

*4.3. Individual fingerprinting with GEFF is potentially task-independent*

This work provides strong evidence to suggest that GEFF makes individual fingerprinting task independent. When we used two or more tasks (one of those being resting-state) to create an average individual representation, we found that GEFF was able to correctly identify a validation FC ($\geq$ 90%) even when it belonged to a task not included to construct the average individual representation. Assuming the task-independent nature of GEFF, specific FCs with embeddings that fall far away from the average representation of a given subject might indicate suboptimal quality of its estimation. This also suggests that perhaps we are all hardwired in a similar way and that there are only subtle differences in terms of functional reconfiguration when performing any cognitive task[48,58]. Therefore, perhaps it is not the task but the individual wiring of the person that explains maximal inter-subject variability.

It must be emphasized that GEFF was *potentially task-independent* only when one of the tasks used to create the average individual representation was resting-state. When resting-state is not used to create the average, the fingerprinting accuracy drops considerably (by as much as 10% in one case). When resting-state is part of the average, by adding more and more tasks into the average, the fingerprinting accuracy approaches perfection (100%). On the other hand, when resting-state is excluded from the average, we found that even though the fingerprinting accuracy increases with increasing number of tasks in the average, it reaches a plateau at 92%. An average representation created from exclusively non-resting-state tasks is not entirely generalizable to identify resting-state FCs, as mentioned before in *section 4.1*. This suggests that resting-state connectivity captures a fingerprint of an individual which is somewhat orthogonal to other tasks, as described in a little more detail below.

*4.4. Resting-state and one task cover the entire cognitive space for individual fingerprint*

When we used two tasks to create average individual representations (centroids) in GEFF, if one of the two is resting-state, we found that the resultant average representation has a strong individual fingerprint within the same tasks and is also highly generalizable to the external tasks. There were eight tasks implemented and acquired in the HCP dataset, all of them targeting different cognitive capacities as well as neural circuits[5,6], and hence providing with a fair representation of individual's cognitive space[26,27]. Considering the breadth and variety of tasks assessed, our results suggest that one resting-state and one non-resting-state task would potentially be enough to fingerprint an individual anywhere in the cognitive space, i.e. when GEFF is used for these or potentially any other set of fMRI tasks.

As mentioned in the previous section (4.3), an average individual representation created exclusively using non-resting-state tasks, does not fully generalize to the resting-state. But if we use one resting-state and one non-resting-state task, the resultant individual fingerprint is potentially universal across the whole cognitive space. All of this suggests that resting-state and all other tasks form two orthogonal axes of a cognitive space in terms of fingerprinting. This fits well with the idea of an "intrinsic architecture" and a "task-general architecture" proposed by Cole et al[48].

Even though we observed high fingerprinting accuracy by combining any non-resting-state task with resting-state, certain tasks performed better than others. For instance, motor task performed the worst, while relational task performed the best when combined with resting-state. This is in agreement with previous results that show that different tasks seem to possess different levels of individual fingerprint[20,21] and that the individual differentiability that is obtained by combining multiple tasks depends very much on the tasks themselves[57].

Based on these observations, we suggest that when designing an experiment that relies on individual differentiability, the experimenter should acquire one resting-state and one non-resting-state to cover as much individual cognitive space as possible. One could tailor the non-resting-state task to ask a desired question but then combine it with resting-state to extract maximal individual fingerprint.

*4.5. Individual Fingerprinting: GEFF is not affected (adversely) by the sample size*

We observed that GEFF is unaffected by increasing sample size (see Figure S5). While the fingerprinting accuracy generally worsened with original FCs as the sample size increased, with GEFF it did not. This was true for all the different scenarios that we studied. In this work, we only go as high as 80 subjects, but if we extrapolate the observed trends to larger datasets (say, consisting of 500 subjects), we expect that individual fingerprinting accuracy for original FCs, especially with external tasks, would go down considerably. At the same time, we do not see any evidence for this trend in GEFF where the fingerprinting accuracy is stable across different sample sizes. Interestingly, when resting-state is excluded from the learning stage and identification is performed for external tasks, the accuracy goes up with increasing data size. We do not have a clear explanation for this phenomenon at this point. A larger dataset might help us dig a little deeper into this sample size behavior.

*4.6. Task Fingerprinting: Robust and generalizable representation of a task does not require a larger sample size*

Using original FCs or GEFF, we show that the task identification accuracy levels off around 20 subjects per task to create the average task representation. With merely 20 samples per tasks, we can create a task representation that is highly accurate (91% for Orig FCs and 96% for GEFF) and highly generalizable to external subjects (90% for Orig FCs and 94% for GEFF). GEFF still outperformed Orig FCs for any number of subjects per task, despite the fact that the performance gap for task identification was not as pronounced as the gap for individual identification. Note that when assessing TI for more than 20 subjects per task, GEFF TI continued to rise, reaching 98% with 80 subjects per task, while Orig FCs TI did not exhibit improvement.

### *4.7. Task Identification is high for all tasks with GEFF*

With only 20 subjects per task to create the average task representation, GEFF was able to identify all eight tasks with an average accuracy of 94.8%, which is comparable to the 93.7% accuracy achieved by a deep learning framework[33]. All the tasks had identification accuracies above 90%, except emotion (86%). Even for external subjects, the average accuracy was 92.4%. Although, we should highlight that in Wang et al.[33], the sample size is much larger ($N = 1034$) than in this study ($N = 100$). We would emphasize again that GEFF was tested here with only eight tasks but we show that this framework has the potential to be universal in decoding large number of cognitive states simultaneously. Thus, GEFF could be employed to track a dynamically changing mental state with high accuracy in a relatively straightforward manner. In addition, using dynamic FC, we could also use GEFF to create a dynamic eigenspace profile of a subject doing different tasks.

### *4.8. The eigenspace displays task- and subject-dominant regimes*

By characterizing eigenvectors based on their task- and/or subject-specificity, we were able to show that they can be delineated into task- and subject-dominant regimes, simply by ordering them in descending order of explained variance. We observed that the first 10 eigenvectors which explained around 80% of the variance in the data, were highly task-dominant, while there was a second wave of eigenvectors from $10 - 300$ that were subject-dominant. Interestingly, most of the eigenvectors were neither task- nor subject-specific.

We should emphasize here that this organization of eigenvectors into specificity regimes was not intuitive to us beforehand. The task- and subject-specificity could easily have been spuriously distributed across the spectrum or there could easily have been no task- or subject-dominant eigenvectors. The fact that by simply ordering eigenvectors in descending order of their explained variance, they are delineates into task- and subject-dominant regimes is an interesting phenomenon.

### *4.9. Future work*

We aim to reproduce these findings with a larger sample size to estimate the effects of increasing number of subjects and tasks on the robustness and task-independence of GEFF. We can potentially also use GEFF with dynamic FCs and create dynamic eigenspace profiles of individuals to see if those profiles provide any additional information about the individual and how that individual reconfigures with changing mental states within a task. We also need to test this framework with different parcellation sizes. GEFF could be used to track disease progression over time and lead to more personalized medicine. In addition, GEFF could be applied to effective connectivity data in much the same as function connectivity data.

*4.10. Limitations*

GEFF was tested with a relatively modest sample size of 100 subjects, although we would like to test this framework with larger datasets. In addition, we only used one parcellation and it has been shown that parcellation size has an effect on the FC fingerprint. New FCs cannot be added dynamically to the dataset with GEFF, as it requires a group-level decomposition to create an eigenspace. So, every time a new FC is added to the dataset, a reconstruction of the eigenspace and a subsequent updated projection of all the data is needed. Also, since whole FCs are embedded as points in a high-dimensional eigenspace, we cannot discern the contribution of individual brain regions to the identification accuracy. The cognitive space of the subjects was explored through the eight available fMRI tasks in the HCP dataset. Datasets with even more fMRI tasks will allow better exploration of subject and task fingerprints within the GEFF framework.

## 5. CONCLUSION

In summary, we propose a graph embedding framework, i.e. GEFF, that is extremely accurate in comparing functional connectomes. We demonstrate this by showing very high subject- and task-identification accuracies using the HCP 100 unrelated subjects dataset. We also show that GEFF is potentially task-independent for subject-identification and subject-independent for task-identification. In other words, the average representation created by GEFF for a subject or task is highly generalizable to external data. In addition, we show that eigenvectors can be characterized as task- or subject-dominant, which provides a deeper insight into the extent of variance of functional connectivity across individuals and cognitive states. GEFF is a robust and potentially universal identification framework that can serve as a potential benchmark for FC fingerprinting and as an exploratory tool to track the cognitive dynamics in an individual.

## ACKNOWLEDGEMENTS

Data were provided [in part] by the Human Connectome Project, WU-Minn Consortium (Principle Investigators: David Van Essen and Kamil Ugurbil; 1U54MH091657) funded by the 16 NIH Institutes and Centers that support the NIH Blueprint for Neuroscience Research; and by the McDonnell Center for Systems Neuroscience at Washington University.

## FUNDING INFORMATION

Authors acknowledge financial support from NIH R01EB022574 (JG), NIH R01MH108467 (JG and JH), Indiana Alcohol Research Center P60AA07611 (JG), and Purdue Discovery Park Data Science Award "Fingerprints of the Human Brain: A Data Science Perspective" (JG).

## Across-Tasks SI rates using > 2 tasks in the Learning Stage: LS$^{(i)}$

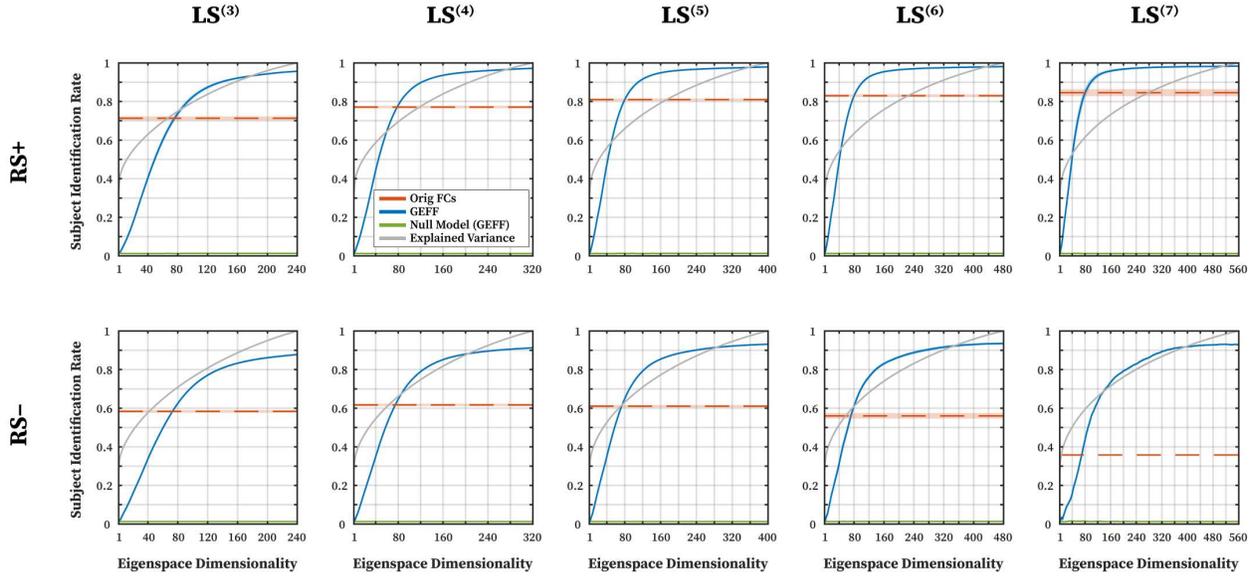

**Figure S1:** Subject Identification (SI) rates for Across-Tasks using more than 2 learning tasks: LS$^{(i)}$, $i = 3, \ldots, 7$. Top row shows the SI curves when one of the learning tasks is resting-state (RS+), while the bottom row shows the curves when resting-state is excluded from the learning tasks (RS−). Columns are tagged by number of tasks ($3 \ldots 7$) used in the learning stage to create the eigenspace. The solid lines show the mean SI rate across all learning tasks' permutations and the shaded regions show the standard error of mean (SEM).

# Validation Stage 1 (Within Subjects)

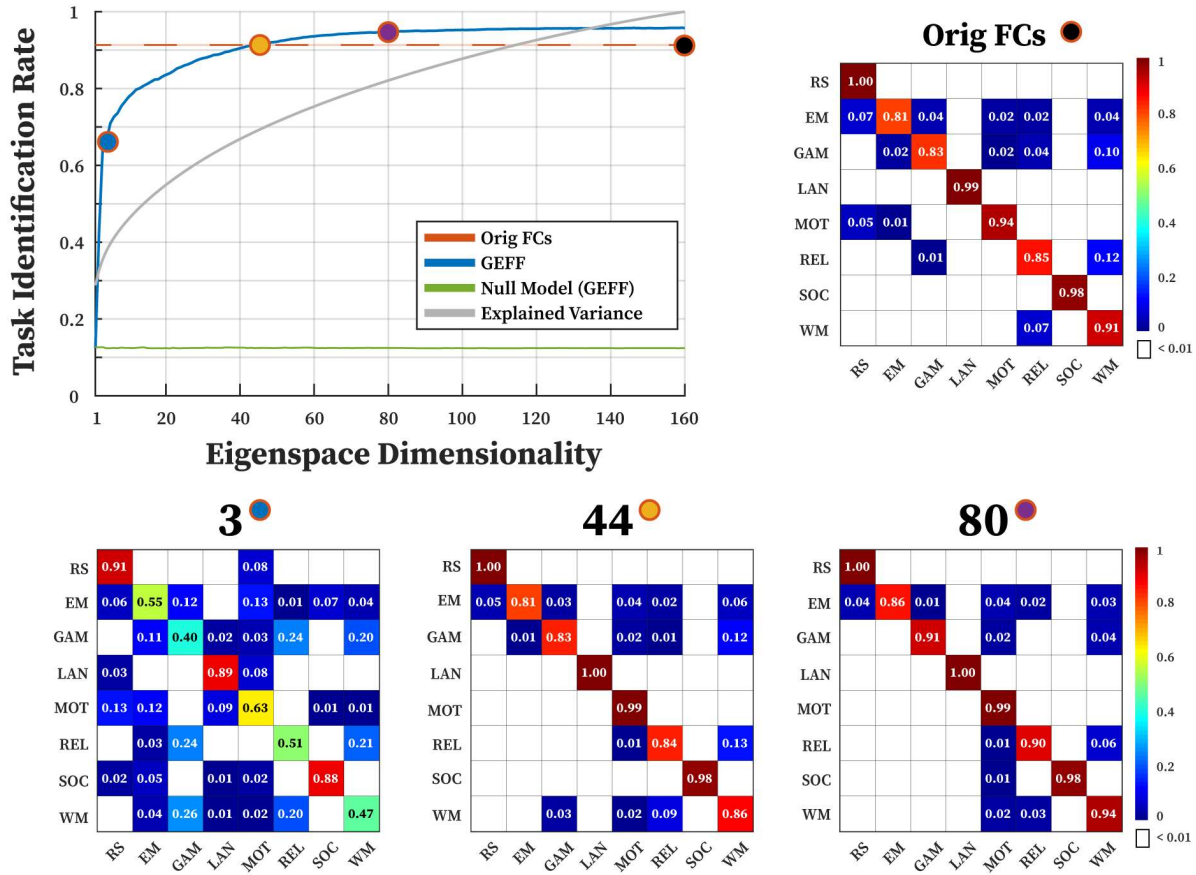

**Figure S2: Task Identification (TI) rates with 20 subjects per task in the Learning stage dataset using Orig FCs and GEFF for *Validation Stage 1*.** The top-left panel shows the TI rate curve with increasing eigenspace dimensionality when the *validation* dataset contains new FCs from the same subjects as the Learning dataset i.e. *Validation Stage 1*. The solid lines show the mean TI rate across bootstrap resamples and the shaded regions show the standard error of mean (SEM). The top-right panel is the confusion matrix describing the performance of TI process using Orig FCs. The main diagonal shows the fraction of validation FCs that were identified correctly for each task. Off diagonal elements show the fraction of FCs which were mislabeled as other tasks. Confusion matrices in the bottom panels correspond to GEFF with eigenspace dimensionality of 3, 44 and 80 as shown above the matrices and highlighted on the TI curve in the top-left panel.

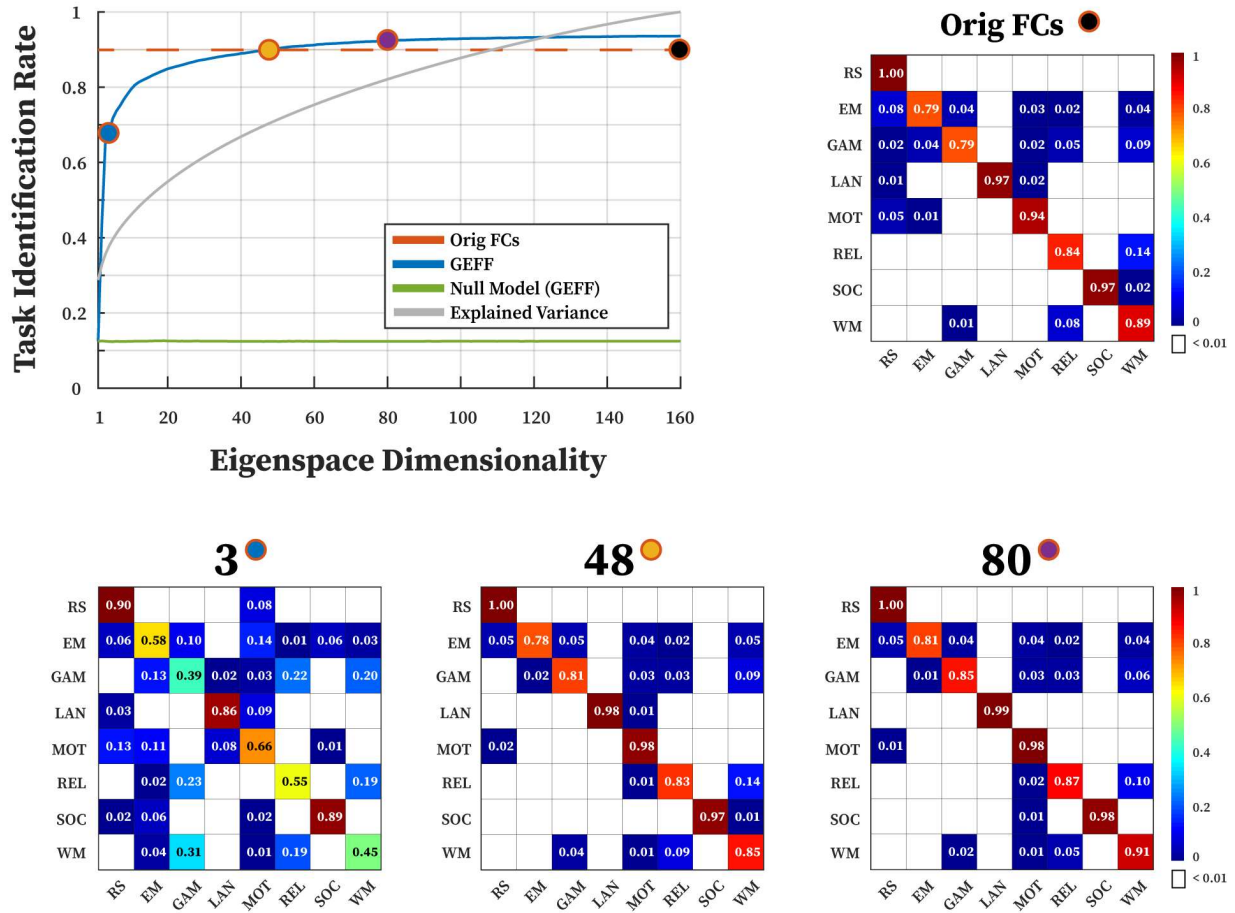

**Figure S3: Task Identification (TI) rates with 20 subjects per task in the Learning stage dataset using Orig FCs and GEFF for *Validation Stage 2*.** The top-left panel shows the TI rate curve with increasing eigenspace dimensionality when the *validation* dataset contains new FCs from the subjects *not* included in the Learning dataset i.e. *Validation Stage 2*. The solid lines show the mean TI rate across bootstrap resamples and the shaded regions show the standard error of mean (SEM). The top-right panel is the confusion matrix describing the performance of TI process using Orig FCs. The main diagonal shows the fraction of validation FCs that were identified correctly for each task. Off diagonal elements show the fraction of FCs which were mislabeled as other tasks. Confusion matrices in the bottom panels correspond to GEFF with eigenspace dimensionality of 3, 48 and 80 as shown above the matrices and highlighted on the TI curve in the top-left panel.

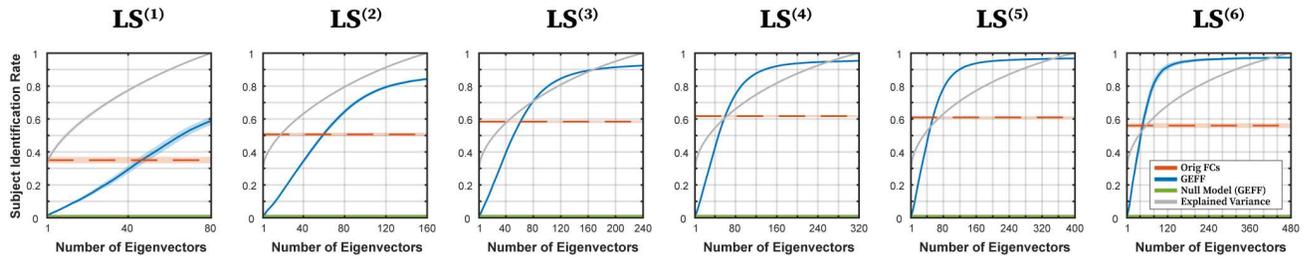

**Figure S4:** Subject Identification rates for Across-Tasks when resting-state is excluded from both Learning and Validation datasets for Orig FCs and GEFF.

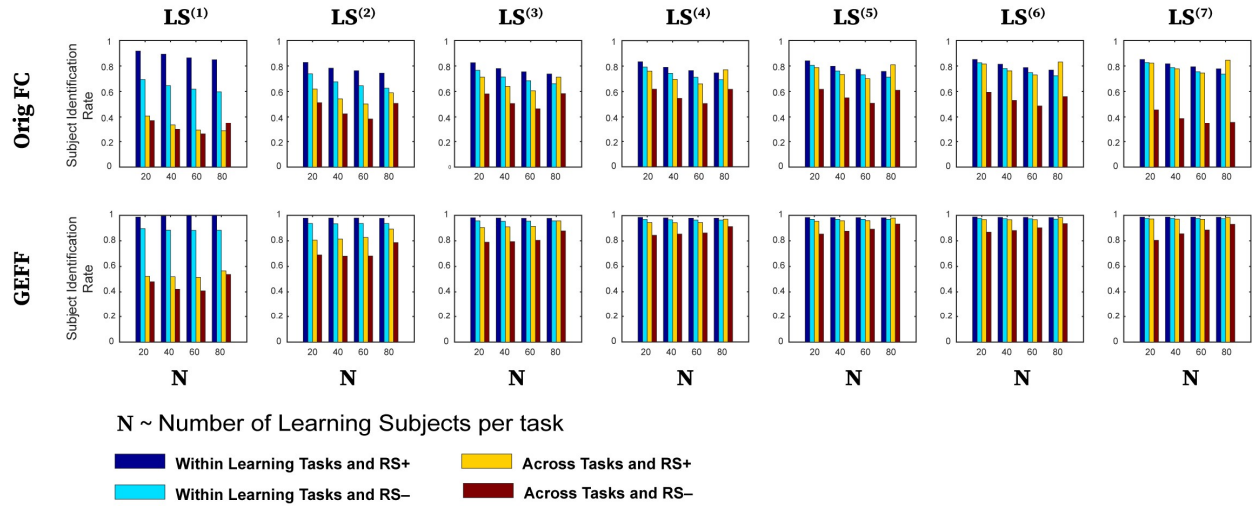

**Figure S5**: Subject Identification (SI) rates for Within-Learning-Tasks and Across-Tasks with increasing number of learning subjects per task.